\documentclass[fleqn,twoside]{article}
\usepackage{espcrc2}
\usepackage{amsmath}

\usepackage{graphicx}

\usepackage[figuresright]{rotating}

\newcommand{\Dslash}{\relax{\kern+.25em / \kern-.70em D}}

\newcommand{\be}{\begin{equation}}
\newcommand{\ee}{\end{equation}}
\newcommand{\bd}{\begin{displaymath}}
\newcommand{\ed}{\end{displaymath}}
\newcommand{\bea}{\begin{eqnarray}}
\newcommand{\eea}{\end{eqnarray}}
\newcommand{\ba}{\begin{array}}
\newcommand{\ea}{\end{array}}

\newcommand{\Id}{\mathbf{1}}
\newcommand{\Real}{\relax{\mathsf{\Gamma\kern-.35em R}}}
\newcommand{\Int}{\relax{\mathsf{Z\kern-.40em Z}}}

\newcommand{\half}{{\scriptstyle{{1\over 2}}}}

\newcommand{\cH}{{\cal H}}

\newcommand{\cL}{{\cal L}}

\newcommand{\cO}{{\cal O}}

\title{Twisted mass QCD and the $\Delta I = 1/2$ rule\thanks{Work supported
in part by the European Community Human Potential
Programme under contract HPRN-CT-2000-00145, Hadrons/Lattice QCD}}

\author{C.~Pena\address[RM2]{I.N.F.N., Sezione di Roma 2,
        c/o Dipartimento di Fisica, Universit\`a di Roma ``Tor Vergata'',\\ 
        Via della Ricerca Scientifica 1, I-00133 Rome, Italy},
        S.~Sint\address{CERN, Theory Division, CH-1211 Geneva 23,
        Switzerland}
        and
        A.~Vladikas\addressmark[RM2]\thanks{Talk presented by A.~Vladikas.}}

\begin{document}

\begin{abstract}
We show that the application of twisted mass QCD (tmQCD) with four (Wilson)
quark
flavours to the computation of lattice weak matrix elements relevant to $\Delta I = 1/2$
transitions has important advantages: the renormalisation of $K\to\pi$ matrix elements
does not require the subtraction of other dimension six operators, the divergence arising
from the subtraction of lower dimensional operators is softened by one power of the
lattice spacing and quenched simulations do not suffer from exceptional configurations
at small pion mass. This last feature is also retained in the tmQCD computation of $K\to\pi\pi$
matrix elements, which, as far as renormalisation and power subtractions are concerned,
has properties analogous to the standard Wilson case.
\vspace{1pc}
\end{abstract}
\maketitle
\section{Introduction}
At scales well below $M_W$, but above the charm quark mass, the effective weak Hamiltonian
for $CP$-conserving, $\Delta S=1$ decays can be written as:
\begin{align}
\cH_{eff} = V_{ud}&V^*_{us}\frac{G_F}{\sqrt{2}} [C_+(\mu/M_W)O_R^+(\mu)
\nonumber \\
&+ C_-(\mu/M_W)O_R^-(\mu) ] \ ,
\label{H_eff}
\end{align}
where  $O_R^{\pm}(\mu)$ are the dimension-6 four-fermion operators
\begin{align}
\label{def_Opm}
\relax{\kern-9mm}
O^{\pm} &= (\bar{s}\gamma_{\mu}^L d)(\bar{u}\gamma_{\mu}^L u) \pm (\bar{s}\gamma_{\mu}^L u)(\bar{u}\gamma_{\mu}^L d)
- [u \leftrightarrow c] \nonumber \\
&= O^{\pm}_{VV+AA} - O^{\pm}_{VA+AV} \ 
\end{align}
renormalised at a scale $\mu$ and $\gamma_{\mu}^L = \half\gamma_{\mu}(\Id-\gamma_5)$
(the subscript $R$ indicates renormalised quantities).
Parity ensures that $K^0\to\pi^+\pi^-$ and $K^0\to 0$ matrix elements will receive contributions only from
$O^{\pm}_{VA+AV}$, while $K^+\to\pi^+$ matrix elements will arise only from $O^{\pm}_{VV+AA}$.

Under isospin transformations, the operator $O^-$ is purely $I=1/2$, while $O^+$ exhibits
both $I=1/2$ and $I=3/2$ parts. The $\Delta I=1/2$ rule can be accounted for by
an enhancement of the $K\to\pi\pi$ matrix element of $C_-O^-$ with respect
to that of $C_+O^+$. The contribution to this enhancement coming from the ratio of Wilson
coefficients $C_-/C_+$ is small in typical renormalisation schemes (such as $\overline{\rm MS}$);
thus, the bulk of the enhancement must come from the ratio of the matrix elements of the operators themselves.

\section{Wilson fermion renormalisation}
A serious difficulty arises upon attempting to extract physical
amplitudes directly from Euclidean correlation functions corresponding to matrix elements with more
than one particle in the final state~\cite{Maiani-Testa}. Although it has recently been shown that
the problem can be bypassed~\cite{periMT_1}, there is still scope in adopting the long-standing alternative to a direct
calculation. This consists in using chiral perturbation theory to obtain $K\to\pi\pi$ matrix elements from $K\to\pi$
ones~\cite{Wilson_theor_1}, the latter not being plagued by the presence of final state interactions.

We are then faced with the problem of operator renormalisation and mixing. Besides a logarithmically divergent
renormalisation constant, the operators $O^\pm$ also mix with two dimension-3 operators ($\bar s \gamma_5d$
and $\bar s d$ with coefficients $c^\pm_P$ and $c^\pm_S$ respectively) and two dimension-5 operators
($\bar s \sigma_{\mu\nu} \tilde G_{\mu\nu} d$ and $\bar s \sigma_{\mu\nu} G_{\mu\nu} d$ with coefficients
$d^\pm_{\tilde \sigma}$ and $d^\pm_\sigma$ respectively). In the case of Ginsparg-Wilson fermions, chiral symmetry
ensures that these subtractions are mild, since $c^\pm_{P,S} \sim \cO(1)$ and
$d^\pm_{\tilde \sigma ,\sigma} \sim \cO(a^2)$ \cite{GW_theor}. Once chiral symmetry is lost with Wilson fermions,
renormalisation patterns significantly worsen. Ignoring operators with contributions to
matrix elements which vanish by the equations of motion we now have (on the basis of $CPS$ symmetry)
\begin{align}
\nonumber
(O_R^{\pm})_{VA+AV} =& \, Z^{\pm}_{VA+AV}(g_0^2,a\mu) \Big[O^{\pm}_{VA+AV} \\
+ c^{\pm}_P(g_0^2,& m, a) \,\, \bar s \gamma_5 d \nonumber \\
\label{VAAV_renW}
+ d^{\pm}_{\tilde\sigma}(g_0^2,& m, a)\bar{s}\sigma_{\mu\nu}\tilde F^{\mu\nu} d +
\cdots  \Big] \ , \\
\nonumber
(O_R^{\pm})_{VV+AA} =& Z^{\pm}_{VV+AA}(g_0^2,a\mu) \Big[O^{\pm}_{VV+AA} \\
\nonumber
+ \sum_{k=1}^4 Z^{\pm}_k&(g_0^2) O^{\pm}_k
+ c^{\pm}_S(g_0^2, m, a) \,\, \bar s d \nonumber \\
\label{VVAA_renW}
+ d^{\pm}_{\sigma}(g_0^2,& m, a)\bar{s}\sigma_{\mu\nu}F^{\mu\nu} d +
\cdots  \Big] \ ,
\end{align}
where $k=VV-AA, SS, PP, TT$ (in standard notation) and the ellipses indicate subtractions
which, being $\cO(a)$, are unimportant to the present discussion\footnote[1]{The coefficient
$d_{\tilde \sigma}$ in eq.~(\ref{VAAV_renW}) is also $\cO(a)$ and could have been omitted.}.
Three problems immediately arise. First, the
subtractions of the dimension-3 operators involve a linearly diverging coefficient
in the case of $O_{VA+AV}$ and, even worse, a {\em quadratically} diverging
coefficient for $O_{VV+AA}$:
\bea
c^\pm_P &\sim& \frac{1}{a} \,\, (m_c - m_u) \,\, (m_s - m_d) \ , \nonumber \\
c^\pm_S &\sim& \frac{1}{a^2} \,\, (m_c - m_u) \ .
\label{Wps}
\eea
A second, milder shortcoming is the mixing of the parity-even operator $O^\pm_{VV+AA}$ with four other
dimension-6 operators (just as in the more familiar case of $B_K$). Finally, Wilson fermions in the quenched
approximation are plagued by exceptional configurations. In ref.~\cite{tmQCD_1} it has been shown
that the tmQCD formulation of Wilson quarks does not suffer from the last two problems. Here we will demonstrate
that the implementation of tmQCD can also reduce the quadratic divergence of $K\to\pi$ matrix elements to a linear one.

\section{tmQCD with four quark flavours}
For a recent review of lattice tmQCD see ref.~\cite{Frezz} and references therein. Here we
extend the formulation to four quark flavours. The action is given by
\bea
\cL &=& \bar{\psi}_l (\Dslash+m_l+i\mu_l \gamma_5) \psi_l \nonumber \\
&+& \bar{\psi}_h (\Dslash+m_h +i\mu_h \gamma_5) \psi_h \ .
\eea
We distinguish a light quark doublet $\psi_l^T = (u,d)$ and a heavy one $\psi_h^T = (s,c)$.
The $2 \times 2$ quark mass matrices are $m_{l,h} = {\rm diag}(m_{u,s}~,~m_{d,c})$
and the twisted mass matrices are $\mu_{l,h} = {\rm diag}(\mu_{u,s}~,~\mu_{d,c})$.
For simplicity we impose mass degeneracy in the light sector; i.e. $m_u=m_d$ and $\mu_u = - \mu_d$.
Two twist angles are then defined through ratios of {\it renormalised} mass parameters:
\begin{gather}
\relax{\kern-8mm}
\tan \alpha = \frac{\mu_{R,u}}{m_{R,u}}~~,~~
\tan \beta  = \frac{\mu_{R,s}}{m_{R,s}} = - \frac{\mu_{R,c}}{m_{R,c}} \ .
\end{gather}

The equivalence of tmQCD and standard QCD is formally established through the axial field rotations \cite{tmQCD_1,Frezz}
\begin{align}
\psi_l &~\to~ \exp \Big[i \alpha \gamma_5\frac{\tau^3}{2} \Big] \psi_l \ ,
\nonumber \\
\psi_h &~\to~ \exp \Big[i \beta \gamma_5\frac{\tau^3}{2}\Big] \psi_h \ ,
\end{align}
(and similarly for $\bar \psi_{l,h}$).

For the $K \to \pi$ weak matrix element a convenient choice is given by $\alpha = \beta = \pi/2$.
Operators are then related as follows
\begin{align}
\big[ P_\pi \big]^{\rm tmQCD} &=~  \big[ P_\pi \big]^{\rm QCD} \ , \nonumber \\
\big[ S_K \big]^{\rm tmQCD} &=~ -i \big[ P_K \big]^{\rm QCD} \ , \nonumber \\
\big[ O^{\pm}_{VA+AV} \big]^{\rm tmQCD} &=~ i \big[ O^{\pm}_{VV+AA} \big ]^{\rm QCD} \ ,
\end{align}
where $P_\pi \equiv \bar d \gamma_5 u$, $S_K \equiv \bar u  s$ and $ P_K \equiv \bar u  \gamma_5 s$.
The following equation between tmQCD and standard QCD {\it renormalised} correlation functions
(at unequal space-time arguments) is thus true up to discretisation effects:
\bea
\langle P_\pi(x) O^{\pm}_{VA+AV}(0)  S_K(y) \rangle_{\rm tmQCD} = \nonumber \\
\langle P_\pi(x) O^{\pm}_{VV+AA}(0) P_K(y) \rangle_{\rm QCD} \ .
\eea
In the asymptotic limit the r.h.s. yields $\langle \pi \vert O^{\pm}_{VV+AA} \vert K \rangle_{\rm QCD}$;
this matrix element can thus also be computed in tmQCD (the l.h.s).
We shall show below that in the latter formalism the renormalisation properties
of the four-fermion operator are much more convenient.

For the $K \to \pi \pi$ matrix element a convenient choice of twist angles is $\alpha = -\beta = \pi/2$.
The equation between the corresponding {\it renormalised} correlation
functions (at unequal space-time arguments)
is then
\begin{align}
\relax{\kern-15mm}
\langle & P_\pi(x) P^\dagger_\pi(y) O^{\pm}_{VA+AV}(0)  S_K(z) \rangle_{\rm tmQCD}  \nonumber \\
& = i \langle P_\pi(x) P^\dagger_\pi(y) O^{\pm}_{VA+AV}(0) P_K(z) \rangle_{\rm QCD} \ ,
\end{align}
where now $P_K \equiv \bar d  \gamma_5 s,~S_K \equiv \bar d s$.
In this case there are no advantages to be gained in the tmQCD computation as far as 
operator renormalisation properties are
concerned. However, in the quenched approximation, the tmQCD computation is free of exceptional configurations.
This is an important advantage close to the chiral limit.

\section{tmQCD renormalisation}
For the ``convenient'' choices of twist angles (i.e. $\alpha = \pi/2, \beta = \pm \pi/2$) and using the
discrete symmetries of the tmQCD action (cf.~\cite{tmQCD_1,Frezz}) we find the following renormalisation pattern
for the operators $O^{\pm}_{VA+AV}$
\begin{align}
(O_R)^{\pm}_{VA+AV} =&~ Z_{VA+AV} \Big[O^{\pm}_{VA+AV} \nonumber \\
    + c^\pm_P &\,\, \bar{s} \gamma_5 d 
    + c^\pm_S \,\, \bar s d + \,\,\, \cdots \Big] \ ,
\end{align}
where the leading behaviour of the coefficients (in the chiral expansion) is
\begin{align}
c^\pm_P ~=~& \frac{1}{a} \,\, F^\pm_P(g_0^2) \,\, (\mu_c - \mu_u) \,\, (\mu_s - \mu_d) \ , \nonumber \\
c^\pm_S ~=~& \,\, F^\pm_S(g_0^2) \,\, (\mu_c^2 - \mu_u^2) \,\, (\mu_s - \mu_d) \nonumber \\
    +& \, G^\pm_S(g_0^2)\,\, (\mu_c - \mu_u) \,\, (\mu_s^2 - \mu_d^2) \ .
\end{align}
Thus, we now have a linear divergence in $c^\pm_P$ (while $c^\pm_S$ contains two terms, both
of $\cO(1)$). This situation compares favourably to the standard QCD case, characterized by
a quadratic divergence ($c^\pm_S \sim 1/a^2$; cf. eq.~(\ref{Wps})). Moreover, there are no
dimension-6 operators to be subtracted in the tmQCD case (in standard QCD there are four
such subtractions; cf. eq.~(\ref{VVAA_renW})).

In order to determine the linearly divergent coefficient  $c^\pm_P$ we must resort to parity
restoration in the continuum limit. Parity is broken by the lattice (Wilson) tmQCD but is
recovered after renormalisation. In particular for the twist angle values of $\alpha = \beta = \pi/2$,
parity transformations in the continuum limit assume the form
\begin{align}
\relax{\kern-6mm} u(x) \to i \gamma_0 \gamma_5 \, u(\tilde x)  &\quad d(x) \to - i \gamma_0 \gamma_5 \, d(\tilde x) \ ,\nonumber \\
\relax{\kern-6mm} s(x) \to i \gamma_0 \gamma_5 \, s(\tilde x)  &\quad c(x) \to - i \gamma_0 \gamma_5 \, c(\tilde x) \ ,
\end{align}
with $\tilde x = (x_0,-\mathbf{x})$ and similarly for the antiquarks. Thus $O^\pm_{VA+AV}$ is a positive parity eigenstate whereas
$P_{ds} \equiv \bar d \gamma_5 s$ is a negative parity eigenstate. This implies that for $x \ne 0$ we have
\bea
\langle (O_R)^\pm_{VA+AV}(0) \,\, (P_R)_{ds}(x) \rangle_{\rm tmQCD} = 0 \ .
\eea
Expressing the above in terms of bare correlations yields
\bea
\langle O^\pm_{VA+AV} \, P_{ds} \rangle &+& c^\pm_P \langle P_{sd} P_{ds} \rangle \nonumber \\
&+& c^\pm_S \langle S_{sd} P_{ds} \rangle = 0 \ .
\label{eq:tmrc}
\eea
Analogous parity arguments can be used to show that the last term of the above expression is a
lattice artifact (i.e. $c_S^\pm \langle S_{sd} P_{ds} \rangle = \cO(a)$) and may be dropped.
Thus, eq.~(\ref{eq:tmrc}) can be solved to determine $c_P^\pm$.

\section*{Acknowledgements}
We acknowledge useful discussions with R.~Frezzotti and G.C.~Rossi.


\begin{thebibliography}{9}
\bibitem{Maiani-Testa}
L.~Maiani and M.~Testa, Phys. Lett. B245 (1990) 585.
\bibitem{periMT_1}
L.~Lellouch and M.~L\"uscher, Commun. Math. Phys. 219 (2001) 31,\\
C.J.D.~Lin~et~al.,~Nucl.~Phys.~B619~(2001)~467.
\bibitem{Wilson_theor_1}
L.~Maiani et al., Nucl. Phys. B289 (1987) 505;\\
C.~Dawson~et~al.,~Nucl.~Phys.~B514~(1998)~313.
\bibitem{GW_theor}
S.~Capitani and L.~Giusti, Phys. Rev. D64 (2001) 014506.
\bibitem{tmQCD_1}
R.~Frezzotti et al., JHEP 08 (2001) 058.
\bibitem{Frezz}
R.~Frezzotti, this conference.
\end{thebibliography}
\end{document}